%% file: article.tex
\documentclass[letter]{aa} 

\usepackage{graphicx}
\usepackage{txfonts}
\usepackage{hyperref}
\usepackage{bm}
\usepackage{pdflscape}
\usepackage{physics}
\include{command}

\usepackage{xcolor}

\usepackage{hyphenat}

\begin{document} 

\title{A common law for the differential rotation of planets and stars}

\titlerunning{A common law for the differential rotation of planets and stars}

\author{
        Vincent G. A. Böning\inst{1}
        \and
        Johannes Wicht\inst{1}
}

\institute{
        {Max-Planck-Institut f\"ur Sonnensystemforschung, Justus-von-Liebig-Weg 3, 37077 G\"ottingen, Germany}\\
        \email{boening@mps.mpg.de}
        }

\authorrunning{Böning \& Wicht}

\date{Received ???; Accepted ???}



\abstract{
All planets and stars rotate. All gas planets in our solar system, the Sun, and many stars show a pattern of east- or westward mean flows. This phenomenon is known as differential rotation in the stellar and as zonal jets in the planetary context.  Observations, laboratory experiments and simulations show that the zonal flow kinetic energy scales like $\ell^{-5}$, where $\ell$ is the spherical harmonic degree (which is effectively a latitudinal wave number). Here, we analyze observation of the Sun, as well as simulations of the dynamics in Saturn and in the outer atmosphere of an ultra-hot Jupiter. While these systems are very different, they all develop strong zonal winds that obey the $\ell^{-5}$ scaling. Our results strongly suggest that there is a simple common mechanism that shapes zonal mean flows in planets and stars independent of 
the flow driving.
}


\maketitle

\section{Introduction}

\label{secIntro}

Differential rotation is a pattern of zonal (eastward or westward) mean flows that circumvent a planet or star. For planets, this phenomenon is also known as zonal flows or zonal jets. It has been observed for all solar system gas planets, for the Sun, and for many stars \citep[e.g.,][]{Howe2009,Benomar2018,Galperin2019ZonalJetsBook}

We are reasonably certain that these mean flows are driven by a statistical correlations of smaller-scale flows known as Reynolds stress. However, it is an open question how exactly the 
energy is fed from the smaller scales to the large scale winds. Several different ideas have been proposed, for instance an inverse cascade from smaller to larger scale eddies and ultimately the winds \citep[e.g.,][]{Rhines1975,Vallis1993}. 
An alternative idea is the mean flow instability suggested by \citet[e.g.,][]{Busse2002}. An initial weak correlation of convective eddies creates a small mean flow. This mean flow will further shear the eddies and thereby reinforce the correlations and mean flow driving. This leads to a runaway effect, where small-scale convection directly drives differential rotation. Recent numerical simulations of fast-rotating deep convection clearly support the picture of small-scale convection directly driving differential rotation via the mean flow instability \citep{Boening2023}. If magnetic fields are strong enough, they also influence the resulting differential rotation profile \citep[e.g.,][]{Hotta2021}. Baroclinically unstable flows may also contribute to shaping differential rotation according to recent evidence \citep{Read2022,Bekki2024}.

A very promising idea is the so-called Rhines scaling \citep{Rhines1975}. \citet{Rhines1975} suggests that zonal jet dynamics on the Gas Giants require that the mean zonal flow speed equals the phase speed of Rossby waves, predicting that the jet width $d$ scales with the characteristic jet velocity $U$ like 
\begin{equation}
\label{d}
    d \approx \sqrt{U/\beta}\;\;,
\end{equation}
where $\beta=2\Omega \sin(\theta)$ with planetary rotation rate $\Omega$ and colatitude $\theta$. 
This explains the jet width at Jupiter and Saturn reasonably well \citep{Heimpel2005,Gastine2014}. 
It also predicts that the kinetic energy of differential rotation should scale as $\ell^{-5}$, where $\ell$ is the harmonic degree 
related to the jet width $d=\pi R/\ell$, where $R$ is radius. 
Velocity $U$ can be calculated from the kinetic energy spectrum per 
degree $\ell$ via 
\begin{equation}
\label{U}
  U = \sqrt{2 \sum_\ell E_Z(\ell,r) } \approx 
      \sqrt{2 \ell E_Z(\ell,r) } \;\; .
\end{equation}
Since the spectrum is very peaked and clearly dominated by 
zonal flows, we can sum over all harmonic contributions. 
Using Equation~\eqref{U} with Equation~\eqref{d} yields the spectrum prediction
\begin{equation}
\label{l-5}
E_Z(\ell,r) = C_Z (\Omega r)^2 \ell^{-5}\;\; ,
\end{equation}
where we have ignored the latitudinal dependence. 

This scaling has indeed been confirmed for the observed wind profiles of Jupiter and Saturn \citep{Galperin2001,Sukoriansky2002}, in two-dimensional simulations \citep[e.g.,][]{Galperin2006,Sukoriansky2007}, in shallow three-dimensional general circulation models of jet formation \citep[][]{Cabanes2020}, and a laboratory experiment \citep[][]{Lemasquerier2022}. \citet{Galperin2008} and \citet{Galperin2019} provide reviews on this topic. 
Recently, it has also been observed in numerical simulations of differential rotation driven in deep domains \citep{Boening2023,Cabanes2024}. The proportionality constant 
$C_Z$ has been shown to be of order one.

Here, we further analyze the kinetic energy spectra of differential rotation in solar observations and in numerical simulations of Saturn's interior and an ultra-hot Jupiter atmosphere. Our aim is to infer whether the -5 power law also holds for these systems. For the Sun, we use differential rotation data from the Helioseismic and Magnetic Imager \citep[HMI,][]{Scherrer2012} onboard the Solar Dynamics observatory \citep[SDO,][]{Pesnell2012}. Regarding simulations, we use snapshots of 
a deep planetary convection model of Saturn \citep{Yadav2020} and of a simulation of the dynamics in the outer atmosphere of an ultra-hot Jupiter \citep[][submitted to A\&A Letters]{Boening2024subcriticalArxiv}.

\begin{figure*}
    \centering
    \includegraphics[width=0.98\linewidth]{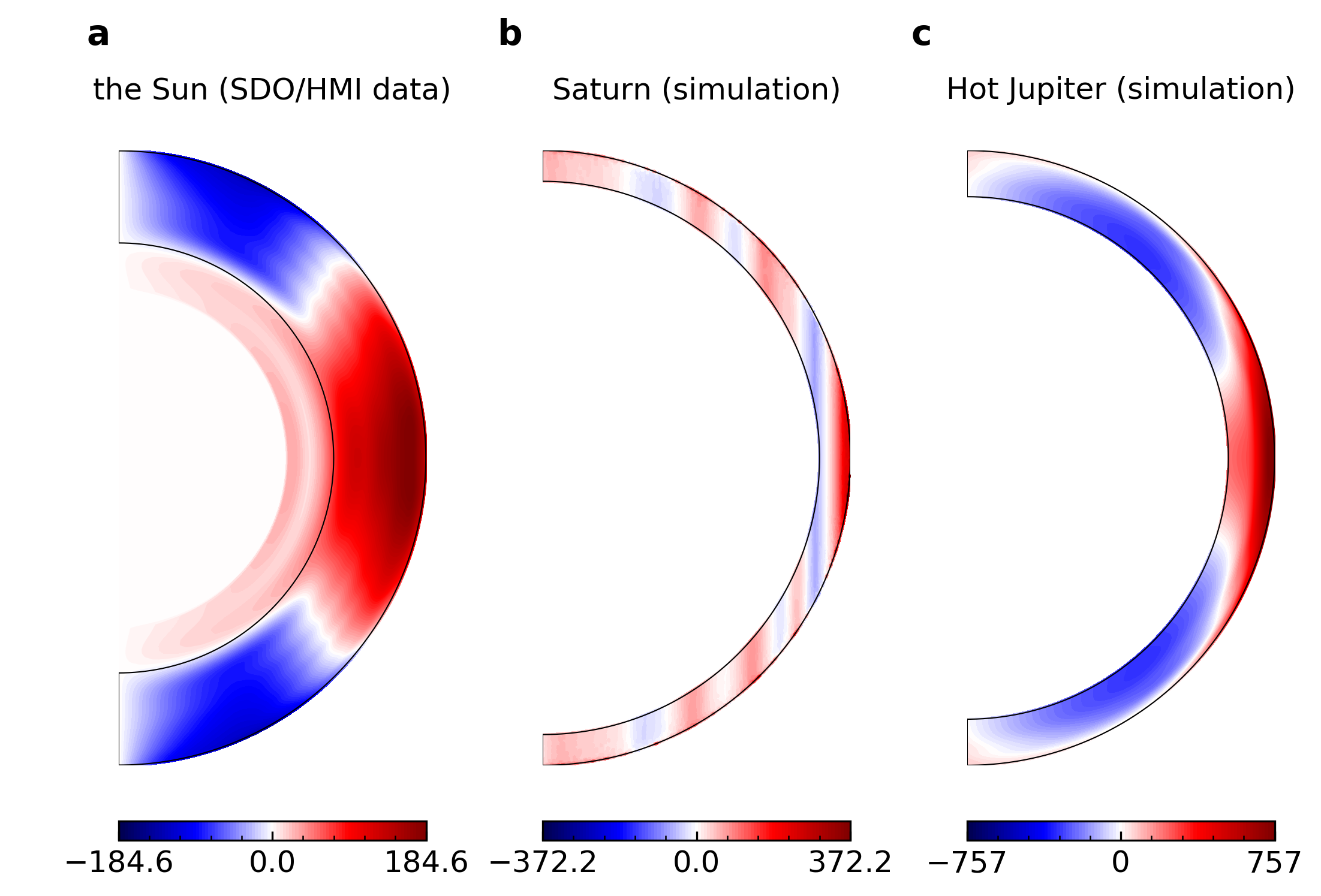}
    \caption{Differential rotation profiles displayed as zonal flow in m s$^{-1}$ from solar SDO/HMI observations (a), in a simulations of Saturn's atmosphere (b), and in a simulation of a an ultra-hot Jupiter atmosphere (c). 
    For the solar observation we show speeds relative to the Carrington rotation rate $\Omega_C=424 \,\rm{nHz}$. 
    Saturn and hot Jupiter simulation were performed with the 
    anelastic version of the MHD code MagIC. MagIC uses a dimensionless formulation and we have rescaled the results to $m/s$ by 
    assuming the planetary rotation rate $\Omega$ and radius $R$. 
    The hot Jupiter has an assumed rotation rate of $2\times 10^{-5}$rad/s and a radius of $84\times10^6$m. 
    Black lines in panel a mark $0.7 \,R_\odot$ and $1.0 \,R_\odot$.}
    \label{fig3profiles}
\end{figure*}

\section{Methods and data}

\label{secMethods}

We analyze three datasets which are either publicly available or have been kindly provided by the original authors. For the Sun, we use data from the Helioseismic and Magnetic Imager \citep[HMI,][]{Scherrer2012} onboard the Solar Dynamics Observatory \citep[SDO,][]{Pesnell2012}. The solar differential rotation data are publicly available online\footnote{We downloaded the  hmi.V\_sht\_2drls time series from \url{http://jsoc.stanford.edu/HMI/Global_products.html}} \citep[see also][]{Larson2018}. The data span a period from 05/2010 until 02/2022 with a gap between 03/2020 and 02/2021 and include eleven 360-day (treated as one-year) averages of the differential rotation rate. We assume a background rotation rate of the Carrington rate, $\Omega_C=424\,\rm{nHz}$. From each one-year average differential rotation rate, we substract the Carrington rate and compute a zonal flow profile using 
\begin{align}
    \bar u_\varphi (r,\theta,\varphi,t) &= r\sin\theta \, \left(\Omega(r,\theta,t) - \Omega_C \right).
\end{align}
Fig. \ref{fig3profiles}a shows the zonal flow profile averaged over 
all available data. The pattern is similar to the well-known differential rotation pattern. The tachocline, which is the transition between the differentially rotating outer envelope and the uniformly rotating interior \citep[e.g.,][]{Miesch2005}, can clearly be seen.

We then perform a spherical harmonic transform of each of the eleven flow profiles $\bar u_\varphi (r,\theta,\varphi,t)$. We compute the spherical harmonic coefficients $u_{\varphi}(\ell,m=0,r,t)$ using the SHTns package \citep{Schaeffer2013} and the normalization implemented in MagIC \citep{Lago2021}. We compute the kinetic energy spectra of the zonal flow for each realization as
\begin{align}
    E_{\rm kin}(\ell,r,t)&= \frac{1}{4\pi} \frac{1}{2} |u_{\varphi}(\ell,m=0,r,t)|^2
\end{align}
and finally compute an eleven-year average spectrum. The resulting kinetic energy spectra are always a function of radius.

\begin{figure*}[h]
    \centering
    \includegraphics[width=0.98\linewidth]{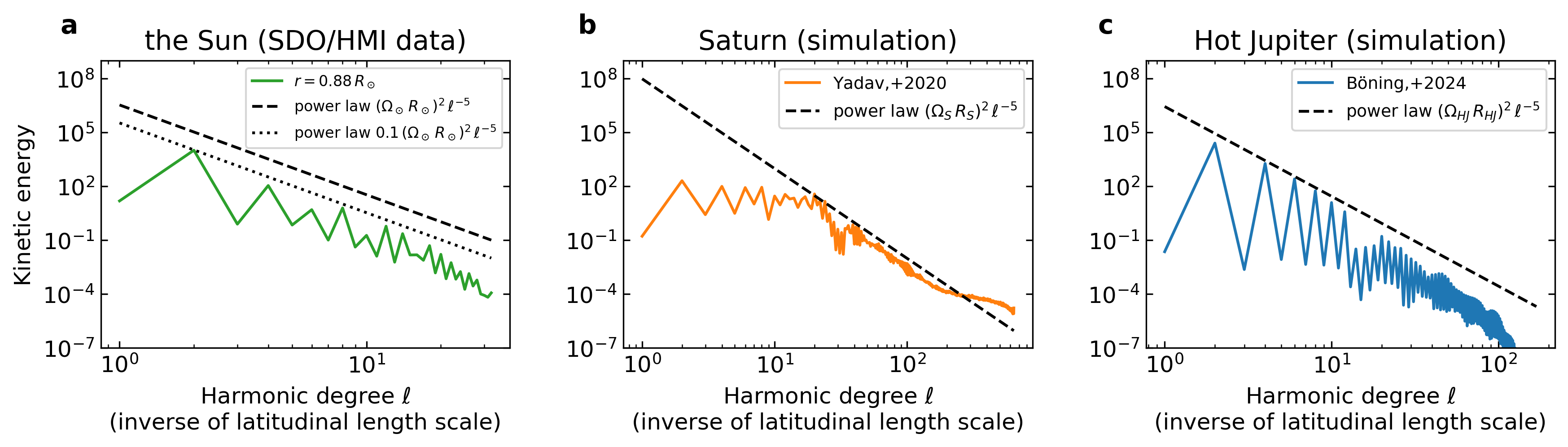}
    \caption{Kinetic energy of differential rotation (zonal mean zonal flows) as a function of spherical harmonic degree 
    for observations of solar differential rotation from SDO/HMI data spanning 05/2010 until 02/2022 (panel a), for jet formation in a deep convection model on Saturn (panel b, original simulation performed by \citealp{Yadav2020}), and for simulations of a hot Jupiter atmosphere (panel c, \citealp{Boening2024subcriticalArxiv}, submitted to A\&A Letters). The dashed line shows the -5 power law predicted by the Rhines scaling and the zonostrophic regime \citep{Rhines1975,Galperin2008}.
    For the scaling of the Saturn and hot Jupiter 
    simulations see caption of Figure \ref{fig3profiles}.}
    \label{figMain}
\end{figure*}

For Saturn, we use a simulation of a deep convecting interior performed by \cite{Yadav2020}, which exhibits Saturn-like jets and a polar hexagon-like feature (see Fig.~\ref{fig3profiles}b).
These authors use the anelastic version of the publicly available code MagIC \citep{Wicht2002} to solve the magnetohydrodynamic equations in a rotating spherical shell that represents the 
outer part of the atmosphere where hydrogen remains molecular and the electrical conductivity is low enough to render magnetic forces secondary. 
The convection is driven by an imposed entropy contrast. For the analysis presented here, we have continued the run by \cite{Yadav2020}. 
The simulation develops strong zonal flows reminiscent of the wind system observed in Saturn's cloud deck. Like on Saturn, a high-latitude 
eastward jet has on polygonal pattern. The jets assume the  
cylindrical geometry that is typical for this kind of models 
\citep{Wicht2020} and reach right through the simulated shell. 
While the jets become roughly stationary in time, they are 
driven by highly time-dependent small-scale convection. 
The solutions is in principal similar to the case analyzed
by \citet{Boening2023} but offers more jets. 

MagIC is also used for simulating the dynamics in the outer 
atmospheres in an ultra-hot Jupiter, which always faces the same side to its host star. In this case, the shell is radially stably stratified but flows are driven by the one-sided radiation. 
The simulation, which is described in more detail in \citet[][submitted to A\&A Letters]{Boening2024subcriticalArxiv}, develops interesting quasi-stationary dynamics dominated by a wide prograde jet around the equator and 
retrograde zonal winds at mid to high latitudes (see Fig.~\ref{fig3profiles}c). 
Contrary to the Saturn simulation, the zonal winds are strongly depth dependent and there is no small-scale time-dependent convection. 
The non-zonal flows are dominated by one large quasi-stationary anticyclonic circulation cell per hemisphere.

For the Saturn and hot Jupiter simulations with MagIC, we used input files representative of the final state of the published work and further integrated the equations using version 6.2, which includes a method to compute the spherical harmonic spectra of the axisymmetric flow components. We use the axisymmetric toroidal kinetic energy spectra, which are identical to kinetic energy spectra of the zonal flow component and divide them by the mass of the simulated shell $M$,
\begin{align}
    E_{\rm kin}(\ell)&= \frac{1}{M} E_{\rm kin,tor,MagIC}(\ell,m=0),
\end{align}
where
\begin{align}
    E_{\rm kin,tor,MagIC}(\ell,m=0)&= \int \frac{1}{2} \,\tilde \rho(r) \,|u_{\varphi}(\ell,m=0)|^2 \,4\pi\,r^2 \,\id r, \label{eqEkinMagic}
\end{align}
and where $\tilde \rho$ is the radially-dependent density of the background state.

We compare the kinetic energy spectra of the observations and simulations to the theoretical spectrum from Equation~\eqref{l-5}. We use the outer radius of a body as using a depth-weighted radius does not make much difference. Since the constant $C_Z$ is expected to be around one, we assume $C_Z=1$ in the following for the simulations \citep[e.g.,][]{Sukoriansky2007}. For the Sun, we also show a curve for $C_Z=0.1$ for comparison. We here use the factor $(\Omega r)^2$ instead of the factor $\beta^2$ with $\beta = \Omega / r$ \citep[e.g.,][]{Sukoriansky2007} because of the dimensionality of the spectrum as defined by spherical harmonics \citep{Lago2021}. 
In both cases, we only compare contributions $\ell \geq 1$, so that the choice of background rotation frame (e.g. the Carrington rate for solar differential rotation) does not severly impact the analysis. The choice of background rotation rate has however a small effect as it enters the energy as a squared quantity.

\section{Results}

\label{secResults}

Figure~\ref{figMain} shows the kinetic energy spectra of differential rotation for the analyzed datasets and a comparison to the predicted scaling. The spectra show a zigzag shape, which results from the dominant equatorial symmetry in the differential rotation profiles. Our analysis therefore focuses on the equatorially symmetric even degree contributions.
As a main result, we find the $-5$ power in major parts of all the analyzed datasets (Fig.~\ref{figMain}).

Figure~\ref{figMain}a shows an average of the 11 individual one-year spectra based on SDO data for a radius of $0.88 \,R_\odot$.
The predicted power law with $C_Z=0.1$ (dotted line) agrees well with the energy at most even harmonics. Energies at $\ell=6$ and $\ell=10$ are below the prediction
and the agreement deteriorates for $\ell>20$. The predicted absolute magnitude of the power law is in rough agreement with observations for $C_Z=0.1$. The usually adopted value of $C_Z=1$ (dashed line) predicts a differential rotation speed faster by about a factor of three.

In Appendix~\ref{appSun}, we explore whether the results for solar differential rotation depend on the averaging procedure or on depth. 
A spectrum for an individual year and the spectrum of the 11-year averaged rotation profile are basically similar to the averaged spectrum
show in Figure~\ref{figMain}a, with minor differences at large $\ell$ that are either due to the long-term time-variation of differential rotation \citep[e.g.][]{Howe2009} or due to noise. 
The agreement with the predicted power law is most convincing in the middle part of the convection zone (Fig.~\ref{fig3layersSun0})). Towards the tachocline ($r=0.78 \,R_\odot$  or $r=0.73 \,R_\odot$ shown in Fig.~\ref{fig3layersSun1}a,b), the spectrum decays more steeply. 
Towards the outer boundary, however, the decay is somewhat shallower 
($r=0.98 \,R_\odot$ shown in Fig.~\ref{fig3layersSun1}c).

Figure~\ref{figMain}b shows a spectrum from the Saturn 
simulation, based on a zonal flow energy integrated over depth 
and over 100 rotations. However, since the zonal flow pattern
is practically stationary, a snapshot analysis yields a similar result. Furthermore, the spectral contributions are integrated over
depth, using the radial dependent (background) density as weight (see Eq.~\ref{eqEkinMagic}). Deeper layers therefore contribute more. 
The agreement with the theoretical spectrum is best in an intermediate range of scales somewhat smaller than the jet scale (equivalent to roughly $\ell=8$). At the largest spatial scales, including the jet scale, the spectrum is roughly flat. Analyses of the observed zonal 
flows in Jupiter and Saturn show that the $\ell^{-5}$ scaling is satisfied in a similar range of scales, that the spectrum also decays significantly less steep than $\ell^{-5}$ at scales larger than the typical 
jet with, but they also appear somewhat less flat than the simulated Saturn spectrum at small $\ell$ \citep{Galperin2001,Sukoriansky2002,Galperin2019}. 
The amplitude is very well
predicted by the theory with a proportionality constant of $C_Z=1$.

Finally, panel c of figure~\ref{figMain} shows the spectrum for 
the hot Jupiter simulation, again depth integrated like 
in the Saturn case. The spectrum represents a snapshot but, even more than in the Saturn simulation, the zonal flows are quasi stationary. 
We find a good agreement with the predicted spectrum for the smallest
even harmonics up to $\ell=12$. For larger $\ell$, the spectrum decays somewhat faster until it drops even more due to the 
applied hyper-diffusion for $\ell>100$.

\section{Discussion}

\label{secDiscussion}

Previous analyses have shown that the $-5$ power law applies to the zonal flows in a wide range of rotating objects, from the zonal winds observed on Jupiter and Saturn \citep{Galperin2001,Sukoriansky2002} over
numerical simulations of two-dimensional turbulence \citep[e.g.,][]{Galperin2006,Sukoriansky2007} or deep three-dimensional convection 
\citep{Boening2023,Cabanes2024} to laboratory experiments \citep{Lemasquerier2022}. 
Here, we show that is also applies to Solar observations, to 
simulations of Saturn's jet dynamics, and to simulations of the 
outer atmosphere of an ultra-hot Jupiter. 
This reinforces the idea that the law captures a universal behavior 
that is largely independent of other dynamical details. 
The hot Jupiter simulation models a stably stratified medium
driven by the one-sided irradiation from the host star and
is thus in stark contrast to the classical deep
convection simulation of Saturn analyzed above. 
Small-scale mechanical driving in the experiments or 
small-scale statistical driving in turbulence simulations also yield 
the same result. 

Remarkable is the fact that the hot Jupiter simulation lacks any 
small-scale dynamics and is practically stationary. There are no traveling Rossby waves. The non-axisymmetric flow is dominated by 
one large stationary anticyclonic circulation cell per hemisphere. 
The zonal wind pattern depends strongly on depth while 
it assumes a (geostrophic) cylindrical symmetry in the 
Saturn simulation and is more or less radially invariant in the
solar observations \citep[e.g., Fig.~19 in][]{Howe2009}. 

What may help towards a more universal understanding of the 
scaling is a somewhat different interpretation of Rhines' idea. 
\cite{Boening2023} have shown that zonal flows are fed from 
all scales without an in-between cascade. The is possible because
the zonal flows themselves cause a deformation of all convective 
features, which ultimately provides the correlation 
necessary to transfer energy into zonal flows. 
This is what \cite{Busse2002} had termed the mean-flow instability, 
a self-enforcing effect that is likely stopped by viscous drag. In this picture, the kinetic energy of zonal flows is provided by the small-scale convection and large-scale flows including large-scale Rossby waves simply act to maintain the equilibrium $\ell^{-5}$ shape.

Rhines' original idea describes a balance between the terms 
corresponding to advection and to Coriolis force in the 
vorticity equation. In the equatorial $\beta$-plane approximation this 
balance reads
\begin{equation}
\label{deform}
    u_x\;\frac{\partial^2}{\partial y^2} U\approx \beta u_x\;\;.
\end{equation}
Here, $y$ decribes the latitudinal and $x$ the longitudinal direction, 
$u_x$ is the non-axisymmetric flow in $x$ direction. 
Assuming a typical latitudinal jet width $d$ for the zonal flow
then yields the expression for the Rhines scale and ultimately 
the $\ell^{-5}$ scaling \citep{Rhines1975}. 
The left-hand side of equation \ref{deform} describes the deformation
of convective (vorticity) feature due to the latitudinal gradients 
in the zonal flow. The scale of the non-axisymmetric flow $u$ does not matter. \citet{Gizon2020} explore how shear in the zonal flow leads to 
modified deformed inertial waves that then yield Reynolds stress. Studying the behaviour of inertial modes at increasing shear rates may help us to understand how the limit in Equation~\ref{deform} comes about. One possibility is that the maximum shear Rossby waves can support is of the order of their azimuthal phase speed, a condition that is equivalent to Equation~\ref{deform}. In a way, our finding is therefore a more universal version of the result of \citet{Bekki2024}, who found that the pole-to-equator amplitude of differential rotation is close to its maximum value.

On the other hand, the $\ell^{-5}$ scaling of zonal flow kinetic energy is also valid for the hot Jupiter simulation, which is quasi-stationary. Although it does not include traveling Rossby waves, it likely includes standing Rossby waves \citep{Showman2011}. The $\ell^{-5}$ scaling may therefore be established by a property of Rossby waves to generate a shear flow of the order of their azimuthal phase speed, which again leads to the $\ell^{-5}$ along the derivation of \citet{Rhines1975}.

Finally, our results open the opportunity to predict latitudinal stellar differential rotation. The only remaining missing link is to understand the signs of the spherical harmonic coefficients in the -5 power law. Commonly, solar-like differential rotation is associated with fast background rotation rates and antisolar differential rotation with slow rotation rates \citep{Gastine2014b}, which would explain the sign of the $\ell=2$ component. If a law for the signs of the other components was found, predicting latitudinal differential rotation profiles would be immediately possible using the -5 power law.

Perhaps the wide-spread applicability of the $\ell^{-5}$ scaling 
simply reflects the universal role of zonal flow shear in  
promoting its own driving or a universal role of Rossby waves in maintaining shear. In any case, our results strongly support the idea that there is a simple common mechanism that shapes zonal mean flows in planets and stars independent of 
the flow driving.



\begin{acknowledgements}
%
We thank P. Read for suggesting to compare the -5 power law to simulations as a reviewer of another paper. We thank T. Gastine, W. Dietrich, U. Christensen, and P. Wulff for discussions. This work was supported by the Deutsche Forschungsgemeinschaft (DFG) in the framework of the priority program SPP 1992 ‘Exploring the Diversity of Extrasolar Planets’. The MAGIC-code is available at an online repository (\url{https://github.com/magic-sph/magic}). This work used NumPy \citep{Oliphant2006,vanderWalt2011}, matplotlib \citep{Hunter2007}, SciPy \citep{Virtanen2020}, and SHTns \citep[][]{Schaeffer2013}.

\end{acknowledgements}


%


\input{article.bbl}
\appendix

\section{Solar differential rotation with different averaging methods and at different radial layers}
\label{appSun}

\begin{figure*}
    \centering
    \includegraphics[width=0.98\linewidth]{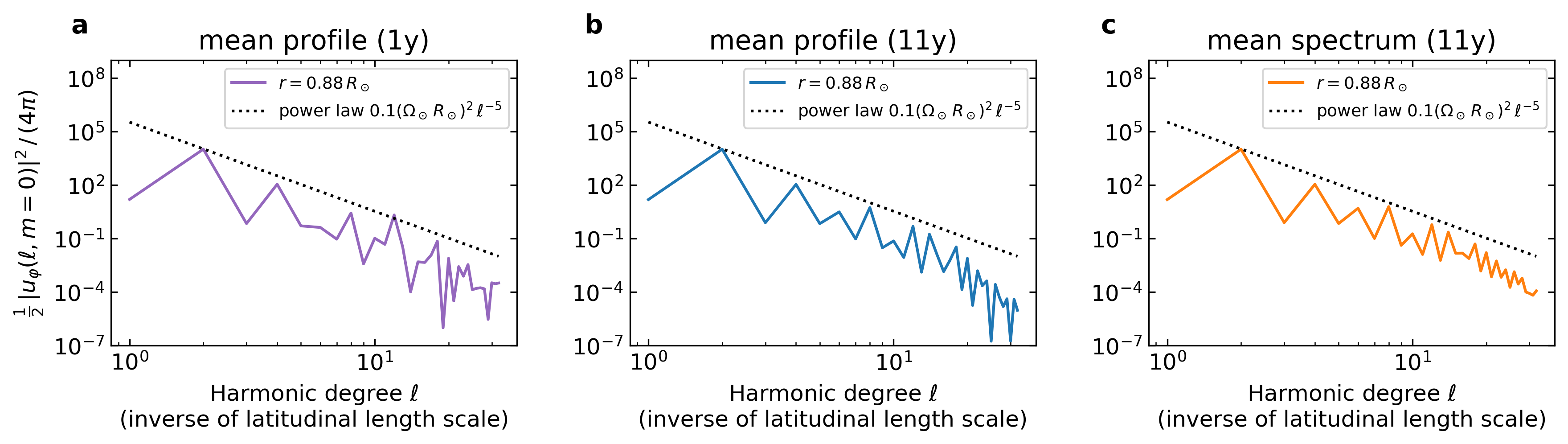}
    \caption{Kinetic energy of differential rotation (zonal mean zonal flows) as a function of the inverse of latitudinal length scale (harmonic degree) for observations of solar differential rotation from SDO/HMI data, comparing detailed in the data analysis procedure. We here show the spectrum obtained from a rotation profile that was averaged over one year of data (panel a, 05/2010 - 04/2011), over 11 years of data (panel b, 05/2010 - 02/2022, with a gap between 03/2020 and 02/2021), and an average spectrum from 11 spectra (panel c) covering the same time period as panel b.}
    \label{fig3methodsSun}
\end{figure*}

Furthermore, we compare different potential averaging methods that can be used to obtain the kinetic energy spectrum of solar differential rotation. Figure~\ref{fig3methodsSun} shows our results. In panel Figure~\ref{fig3methodsSun}a, we show the kinetic energy spectrum obtained from a one-year average rotation profile using a spherical harmonic transform of the zonal flow profile. Figure~\ref{fig3methodsSun}b shows a spectrum that was similarly obtained from a rotation profile that was averaged over the entire dataset used in this study (approx. 11 years of data). Finally, Figure~\ref{fig3methodsSun}c shows the average of the eleven spectra, each obtained for a one-year average in the same way as those shown in Figure~\ref{fig3methodsSun}a. The results agree well, taking the larger noise level in Figure~\ref{fig3methodsSun}a into account. These results further underline the robustness of the observed -5 power law for solar differential rotation. The differences between panels b and c at the smallest scales may be the subject of further study and could be due to the temporal dependence of the rotation profile, which may have lead to a cancellation of smaller-scale features in the 21-year average, or they may be due to different properties of the background noise.

\begin{figure*}
    \centering
    \includegraphics[width=0.98\linewidth]{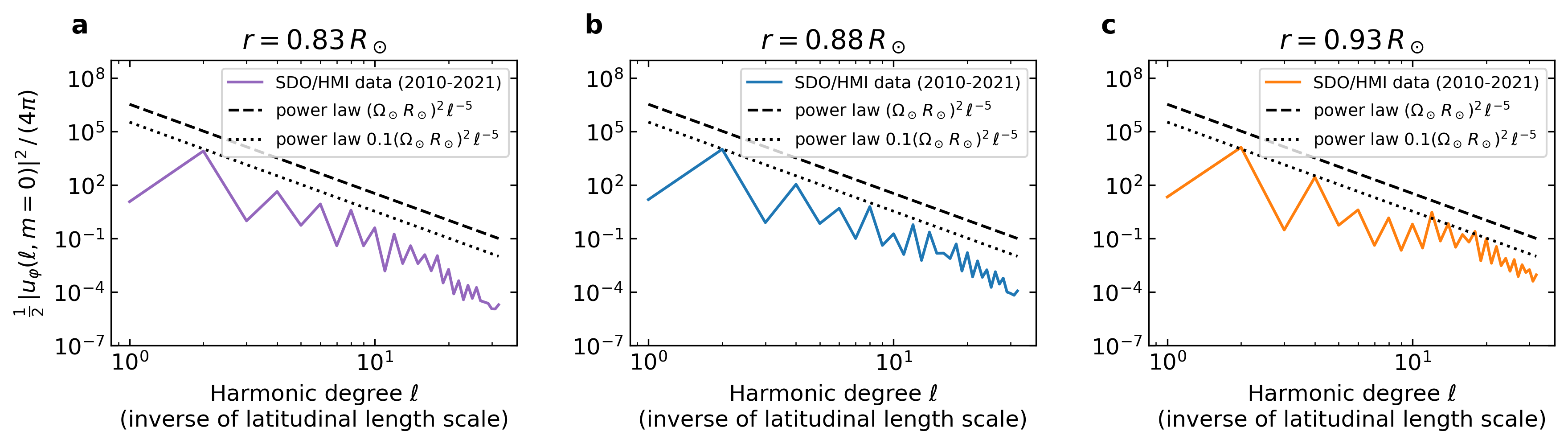}
    \caption{Kinetic energy of differential rotation (zonal mean zonal flows) as a function of the inverse of latitudinal length scale (harmonic degree) for observations of solar differential rotation at different radial layers from SDO/HMI data spanning 05/2010 - 02/2022, with a gap between 03/2020 and 02/2021. The dashed line shows the -5 power law predicted by the Rhines scaling and the zonostrophic regime \citep{Rhines1975,Galperin2019}.}
    \label{fig3layersSun0}
\end{figure*}

\begin{figure*}
    \centering
    \includegraphics[width=0.98\linewidth]{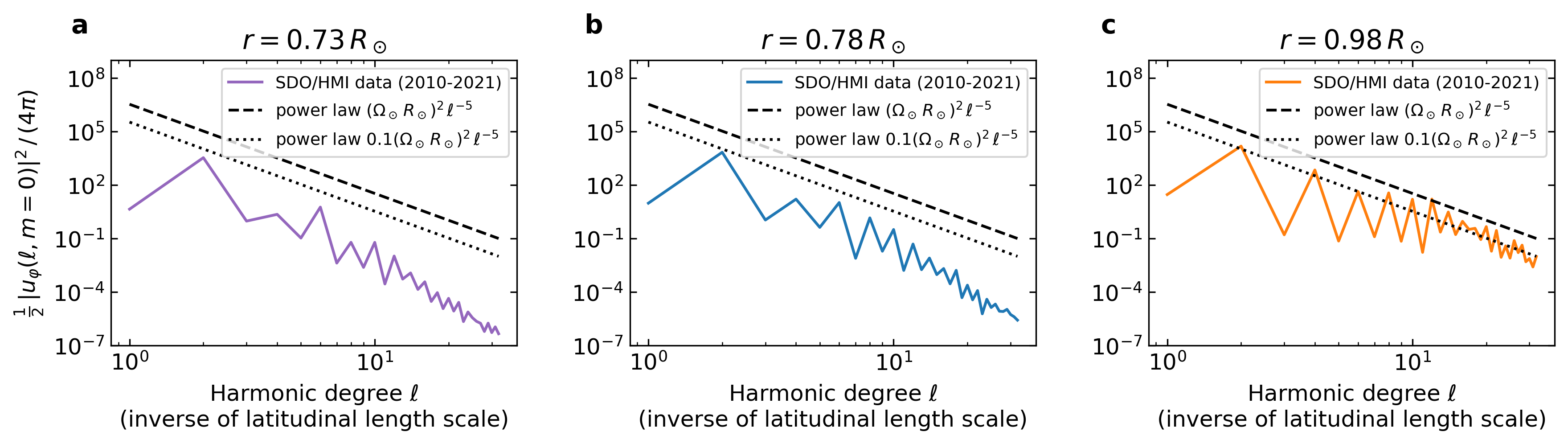}
    \caption{Kinetic energy of differential rotation (zonal mean zonal flows) as a function of the inverse of latitudinal length scale (harmonic degree) for observations of solar differential rotation at different radial layers from SDO/HMI data spanning 05/2010 - 02/2022, with a gap between 03/2020 and 02/2021. The dashed line shows the -5 power law predicted by the Rhines scaling and the zonostrophic regime \citep{Rhines1975,Galperin2019}.}
    \label{fig3layersSun1}
\end{figure*}

Finally, we compare the validity of the -5 power law at different depths in the solar convection zone. Figure~\ref{fig3layersSun0} shows results for the middle and upper middle parts of the convection zone, at $r=0.83 \,R_\odot$, $r=0.88 \,R_\odot$, and $r=0.93 \,R_\odot$. We find that the -5 power law is rather well satisfied in this region, with the best match obtained at $r=0.88 \,R_\odot$. Figure~\ref{fig3layersSun1} shows results for the lower and top part of the convection zone, at $r=0.73 \,R_\odot$, $r=0.78 \,R_\odot$, and $r=0.98 \,R_\odot$. At these levels, the -5 power law is satisfied at the largest scales, with some differences visible at smaller scales. At the topmost layer, these differences may be caused by the presence of strong convective motions that add to the kinetic energy spectrum. The reason for the differences in the deepest layers is unclear to us. One might speculate about different dynamics in this layer either due to a change in stratification, different flow regimes, presence of strong magnetic fields, or generally the effect of neighborhood to the boundary of the convection zone.

\end{document}

%% file: command.tex
\newcommand\id{{\mathrm d}}

\newcommand\Cref{C_{\text{ref}}}
\newcommand\cref{\Cref}